\documentclass[conference]{IEEEtran}
\IEEEoverridecommandlockouts
\usepackage{cite}
\usepackage{hyperref}
\usepackage{amsmath,amssymb,amsfonts}
\usepackage{algorithmic}
\usepackage{graphicx}
\usepackage{textcomp}
\usepackage{xcolor}
\usepackage{caption} 
\usepackage{booktabs}
\usepackage{pgfplots}
\usepackage{pgfplotstable}
\usepackage{multirow}
\usepackage{float}
\restylefloat{table}

\pgfplotsset{compat=1.18}
\def\BibTeX{{\rm B\kern-.05em{\sc i\kern-.025em b}\kern-.08em
    T\kern-.1667em\lower.7ex\hbox{E}\kern-.125emX}}

\newcommand{\IDRAppr}{32\%}

\newcommand{\totaldbmessages}{18{,}797}
\newcommand{\totaldbmessageshundreds}{18{,}700}

\newcommand{\allthreads}{2,638}
\newcommand{\allthreadshundreds}{2,600}
\newcommand{\respondedthreads}{1,285}

\newcommand{\takeoff}{48.7\%}

\newcommand{\totalmessages}{18{,}797} % Total number of messages across all modes
\newcommand{\totalspecialcontentmessages}{14} % Total number of messages with special content
\newcommand{\specialcontentpercent}{0.1\%} % Percentage of messages with special content

\newcommand{\messagesmodeone}{13{,}006} % Messages in Mode 1
\newcommand{\messagesmodetwo}{5{,}791} % Messages in Mode 2

\newcommand{\attackermsgsone}{6{,}289} % Attacker messages in Mode 1
\newcommand{\attackermsgstwo}{3{,}031} % Attacker messages in Mode 2
\newcommand{\attackermsgsall}{9{,}320} % Attacker messages in Mode 2
\newcommand{\defendermsgsone}{6{,}717} % Defender messages in Mode 1
\newcommand{\defendermsgstwo}{2{,}760} % Defender messages in Mode 2

\newcommand{\convsone}{1{,}865} % Conversations in Mode 1
\newcommand{\convstwo}{773} % Conversations in Mode 2

\newcommand{\avgmsgcharone}{342.2} % Average message length in Mode 1
\newcommand{\avgmsgchartwo}{497.7} % Average message length in Mode 2
\newcommand{\avgmsgcharall}{390.0} % Average message length overall
\newcommand{\avgattackerchar}{632.2} % Average attacker message length
\newcommand{\avgdefenderchar}{152.4} % Average defender message length
\newcommand{\durationmodeone}{120} % Days for Mode 1
\newcommand{\durationmodetwo}{34} % Days for Mode 2
\newcommand{\msgsperdaymodeone}{108.4} % Messages/day in Mode 1
\newcommand{\msgsperdaymodetwo}{170.3} % Messages/day in Mode 2
\newcommand{\msgsperconvone}{7.0} % Messages per conversation in Mode 1
\newcommand{\msgsperconvtwo}{6.6} % Messages per conversation in Mode 2

\newcommand{\avglevdistance}{14.4} % Avg Levenshtein distance

\newcommand{\respondedthreadscount}{1{,}285} % Threads with at least one scammer response
\newcommand{\threadsbyallmodeone}{1{,}865} % Mode 1 total threads
\newcommand{\threadsbyallmodetwo}{773} % Mode 2 total threads

\newcommand{\threadsbyvalidmodeone}{891}
\newcommand{\threadsbyvalidmodetwo}{394}
\newcommand{\threadsbyvalidmodeonepct}{47.8\%}
\newcommand{\threadsbyvalidmodetwopct}{51.0\%}

\newcommand{\meanthreadlength}{13.4}
\newcommand{\medianthreadlength}{9.0}
\newcommand{\threadlengthstddev}{14.9}
\newcommand{\threadlengthmin}{2}
\newcommand{\threadlengthmax}{206}

\newcommand{\threadstwotofivepct}{35.1\%}

\newcommand{\threadssixtotenpct}{22.5\%}
\newcommand{\threadseleventotwenty}{295}

\newcommand{\threadshundredplus}{4}

\newcommand{\avgthreadduration}{10.55}
\newcommand{\medianthreadduration}{4.37}

\newcommand{\validthreadsone}{891}
\newcommand{\meanlengthone}{14.2}

\newcommand{\avgdurationone}{12.94}

\newcommand{\validthreadstwo}{394}
\newcommand{\meanlengthtwo}{11.7}

\newcommand{\avgdurationtwo}{5.16}

\newcommand{\longestthreadlen}{206}
\newcommand{\longestthreaddur}{56.07}

\newcommand{\longestduration}{139.01}
\newcommand{\longestdurationmsgcount}{20}

\newcommand{\singleresponsecount}{1{,}170}
\newcommand{\singleresponsepercent}{46\%}

\newcommand{\IDRoverall}{17.66\%}
\newcommand{\IDRconvonly}{31.74\%}

\newcommand{\IDRmodeone}{17.80\%}
\newcommand{\IDRmodetwo}{17.34\%}

\newcommand{\IDRmodeoneconv}{30.91\%}
\newcommand{\IDRmodetwoconv}{34.01\%}

\newcommand{\IDRsuccessthreads}{466}

\newcommand{\harperfectmatch}{1,905}
\newcommand{\humanacceptancerate}{69.02\%}

\begin{document}

\title{"Send to which account?"\\ Evaluation of an LLM-based Scambaiting System}

\author{
Hossein Siadati$^{*}$,
Haadi Jafarian$^{\dagger}$,
Sima Jafarikhah$^{\ddagger}$\\[0.5em]
\begin{tabular}{c}
$^{*}$AI Research, Cybera Global Inc./ UNCW, Wilmington, USA \\
\texttt{s.h.siadaty@gmail.com}
\end{tabular}\\[0.5em]
\begin{tabular}{c}
$^{\dagger}$Department of Computer Science and Engineering, UC Denver, USA \\
\texttt{haadi.jafarian@ucdenver.edu}
\end{tabular}\\[0.5em]
\begin{tabular}{c}
$^{\ddagger}$Department of Computer Science, UNCW, Wilmington, USA \\
\texttt{jafarikhaht@uncw.edu}
\end{tabular}
}

%\author{\IEEEauthorblockN{1\textsuperscript{st} Hossein Siadati}
%\IEEEauthorblockA{\textit{AI Research} \\
%\textit{Cybera Global Inc.}\\
%Wilmington, USA \\
%s.h.siadaty@gmail.com}
%\and
%\IEEEauthorblockN{2\textsuperscript{nd} Haadi Jafarian}
%\IEEEauthorblockA{\textit{Department of Computer Science and Engineering} \\
%\textit{UC Denver}\\
%Denver, USA \\
%haadi.jafarian@ucdenver.edu
%}
%\and
%\IEEEauthorblockN{3\textsuperscript{rd} Sima Jafarikhah}
%\IEEEauthorblockA{\textit{Department of Computer Science } \\
%\textit{UNCW}\\
%Wilmington, USA \\
%jafarikhaht@uncw.edu}

%}

\maketitle

\begin{abstract} Scammers are increasingly harnessing generative AI (GenAI) technologies to produce convincing phishing content at scale, amplifying financial fraud and undermining public trust. While conventional defenses, such as detection algorithms, user training, and reactive takedown efforts remain important, they often fall short in dismantling the infrastructure scammers depend on, including mule bank accounts and cryptocurrency wallets. To bridge this gap, a proactive and emerging strategy involves using conversational honeypots to engage scammers and extract actionable threat intelligence.

This paper presents the first large-scale, real-world evaluation of a scambaiting system powered by large language models (LLMs). Over a five-month deployment, the system initiated over \allthreadshundreds\ engagements with actual scammers, resulting in a dataset of more than \totaldbmessageshundreds\ messages. It achieved an Information Disclosure Rate (IDR) of approximately \IDRAppr, successfully extracting sensitive financial information such as mule accounts. Additionally, the system maintained a Human Acceptance Rate (HAR) of around 70\%, indicating strong alignment between LLM-generated responses and human operator preferences. Alongside these successes, our analysis reveals key operational challenges. In particular, the system struggled with engagement takeoff: only \takeoff\ of scammers responded to the initial seed message sent by defenders. These findings highlight the need for further refinement and provide actionable insights for advancing the design of automated scambaiting systems.
\end{abstract}


% Conference Page:  \url{https://apwg.org/event/ecrime2025/}

\section{Introduction}

According to the Federal Trade Commission (FTC), consumers reported over \$12.5 billion in fraud losses in 2024, a 25\% increase over the prior year. Moreover, 38\% of individuals who reported fraud experienced a financial loss, representing an 11\% rise from 2023. Notably, scams involving payments through bank transfers or cryptocurrency resulted in greater total losses than all other payment methods combined~\cite{FTC2025FraudLosses}. These trends have imposed substantial harm on victims and contributed to a growing erosion of trust in institutions. The widespread adoption of GenAI tools has significantly amplified the phishing threat landscape, adding to the complexity of existing defenses. These tools enable scammers to effortlessly craft highly persuasive phishing messages across multiple languages, personas, and contexts, dramatically lowering the barrier to large-scale manipulation of potential victims~\cite{infosecurity2023_chatgpt_phishing, CNBC2023PhishingAI, gressel2024discussion, ShibliPritomGupta2024}.

A broad spectrum of defenses exists to counter scams, ranging from detection mechanisms to user education initiatives. Among these, detecting and blocking scam messages remains a primary defense strategy; however, it often resembles a game of whack-a-mole, as scammers rapidly adapt their tactics. This dynamic has further intensified with the growing use of AI by both attackers and defenders, each seeking to outmaneuver the other. While tools that help users make informed decisions during suspicious interactions offer some degree of protection, they too are entangled in this adversarial cycle.

A more strategic approach to scam defense targets the core infrastructure enabling scammers’ monetization, such as mule bank accounts and cryptocurrency wallets~\cite{fbiMule, florencio2010phishing}. Identifying and disrupting these financial channels can significantly weaken scam operations, with uncovering this infrastructure as the critical first step.

Banks typically employ a combination of transaction monitoring, such as Anti-Money Laundering (AML) systems, and behavioral analysis to detect mule accounts, especially when indications of account compromise arise. These methods are increasingly augmented by crowdsourced intelligence and threat data. Threat feeds and reports from victims or honeypots can expose mule infrastructure, including bank account numbers, cryptocurrency wallet addresses, and IP addresses.

Traditional intelligence-gathering techniques, such as dark web or black market monitoring, tend to produce broad, non-targeted data and often suffer from latency or incomplete coverage. In contrast, financial institutions actively share intelligence via industry platforms such as FS-ISAC or sector-specific threat watchlists, which can enhance the precision and timeliness of mule account detection.

\textit{Scambaiting} refers to the proactive engagement with scammers by impersonating victims through deceptive engagements, with the goals of wasting the attacker's time, disrupting their operations, and gathering threat intelligence. Scambaiting technique has shown promise in generating timely and actionable intelligence~\cite{siadati2020_framework_attackers_accounts, Siadati2019}. Once a manual and time-consuming task, scambaiting is increasingly becoming automated. Recent research highlights the potential of LLM-based agents to autonomously converse with scammers~\cite{basta2025bot, charnsethikul2025puppeteer}. In this context, a gen model (e.g., ChatGPT) can be prompted to produce responses that convincingly impersonate a potential victim. When carefully constructed, such interactions can generate valuable disclosures from scammers, such as bank account details or cryptocurrency wallet addresses. This information can then be used to proactively intervene in scam operations, for example, by flagging and disabling the identified financial accounts through collaboration and coordination with financial institutions. 

Despite growing interest in LLM-driven security applications, measuring the effectiveness of LLM-powered scambaiting for scam intelligence collection remains largely unexplored. In this paper, we present the first large-scale evaluation of a real-world deployment of such a system.

%\textcolor{red}{needs work}
This paper makes several important contributions to the study of LLM-powered adversarial dialogue systems:

\begin{itemize}

    \item \textbf{Large-scale analysis of real-world scammer interactions.} To our knowledge, this is the first large-scale evaluation of an operational LLM-based scambaiting system engaging real-world scammers over a five-month deployment. The resulting dataset includes \totaldbmessages\ messages across \allthreads\ attacker-defender engagements initiated by human adversaries.
    
   \item \textbf{Comprehensive evaluation framework.} We designed and implemented a set of specialized metrics for assessing automated scambaiting systems. These  include the \textit{Information Disclosure Rate (IDR)}, \textit{Information Disclosure Speed (IDS)}, and \textit{Message Freshness}, which respectively measure the effectiveness, efficiency, and linguistic variability of baiting engagements.

    \item \textbf{Actionable insights for real-world deployment.} Our empirical findings yield practical guidance for deploying LLM-driven proactive security tools. These include engagement death thresholds, engagement design strategies that increase elicitation success, and observations on the impact of human-in-the-loop interventions on responsiveness and outcomes.

\end{itemize}

The remainder of this paper is organized as follows: Section~\ref{sec:background} reviews relevant background and related work in scambaiting and adversarial LLM dialogue. Section~\ref{sec:system} describes our system architecture and deployment methodology. Section~\ref{sec:dataset} details the data collection and annotation process. Section~\ref{sec:evaluation} presents our core evaluation metrics and empirical results, including responsiveness, disclosure patterns, and human-in-the-loop impact. Section~\ref{sec:analysis} presents a comprehensive and in-depth analysis, uncovering key patterns and insights.
Section~\ref{sec:insights} discusses key operational insights and limitations. Finally, Section~\ref{sec:conclusion} concludes with takeaways and future research directions.

 \begin{figure}[t!]
    \centering
    \includegraphics[width=0.4\textwidth]{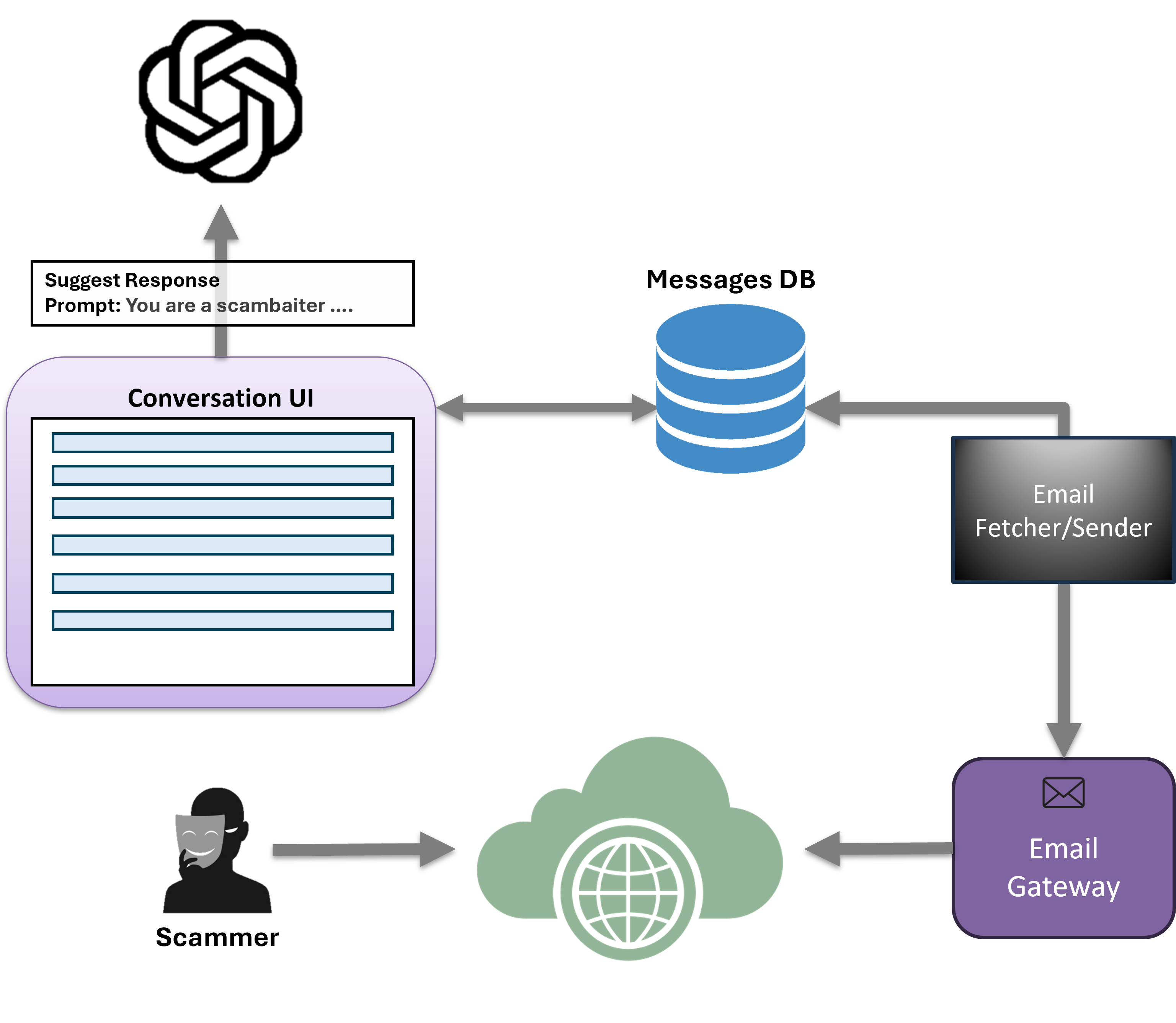}
    \caption{System architecture of the scambaiting platform that collected data for this study. The Conversations UI displays scammer interactions and sends prompts (e.g., “You are a scambaiter...”) to the LLM (ChatGPT) to suggest responses. Approved replies are routed through the email gateway to scammers. All messages are stored in a central Messages DB and managed by an email fetcher/sender service. The system facilitates live engagement with scammers, with support for human-in-the-loop participation and oversight.}
    \label{fig:system-architecture}
\end{figure}

\section{Background and Related Work}\label{sec:background}

\begin{table*}[t!]
\centering
\caption{Summary of Evaluation Criteria in Prior Scambaiting Systems}
\label{tab:evaluation-criteria}
\begin{tabular}{p{2.8cm}|p{3.8cm}|p{2.2cm}|p{5.5cm}}
\hline
\textbf{System / Paper} & \textbf{Success Metric(s)} & \textbf{Type} & \textbf{Description} \\
\hline
\textbf{Puppeteer}~\cite{charnsethikul2025puppeteer} & Information Disclosure Rate & Functional & Percentage of interactions where scammers disclosed sensitive data (e.g., bank details); compared across LLM-only vs. orchestrated setup. \\
\cline{2-4}
& Convincingness & Human-rated & Binary success metric based on whether simulated scammers believed the LLM was a real victim. \\
\cline{2-4}
& Persona Coherence & Heuristic & Consistency of LLM responses with a predefined victim persona across the dialogue. \\
\hline
\textbf{Bot Wars Evolved}~\cite{basta2025bot} & Engagement Length & Structural & Number of turns or duration of scammer-victim dialogue between competing LLMs. \\
\cline{2-4}
& Goal Fulfillment & Functional & Whether each LLM (scammer or victim) achieved its role-specific objective. \\
\cline{2-4}
& Conversational Realism & Human-rated & Assessed by comparing generated dialogues with real-world scambaiting transcripts. \\
\hline
\textbf{Re:Scam (1.0 \& 2.0)}~\cite{netsafe2024,netsafe2024rescam} & Scammer Time Wasted, Engagement Realism, Persona Diversity & Structural, Heuristic & Focused on misleading and prolonging interactions with scammers via scripted replies (1.0) and AI-generated personas (2.0). Success inferred through engagement length, response variation, and anecdotal evidence. \\
\hline
\end{tabular}
\end{table*}

\subsection{LLMs for Scambaiting}
Early scambaiting approaches primarily relied on scripted or rule-based systems. For instance, Netsafe's \emph{Re:Scam}~\cite{netsafe2024, vincent2017rescam} responded to scam emails using fixed conversational templates, while the voice-based “Lenny” chatbot stalled phone scammers using repetitive prerecorded audio clips~\cite{vincent2017rescam}. Although effective at confusing human attackers, these systems suffered from limited adaptability and linguistic variation due to their static designs.

Recent advances in GenAI and LLMs have enabled more flexible and realistic interactions. Netsafe's \emph{Re:Scam 2.0}~\cite{netsafe2024rescam} leveraged LLMs to diversify victim personas and improve conversational coherence. These systems marked a transition from rule-based methods to generative, persona-rich engagements.

Charnsethikul et al.'s \emph{Puppeteer}~\cite{charnsethikul2025puppeteer} combined LLM-generated messages with finite-state dialogue management to simulate victims of Amazon refund scams. In controlled evaluations, it achieved a 68\% Extraction Score for sensitive information (i.e., shipping address in Amazon customer service scam), outperforming unstructured LLMs (50\%), and demonstrating the benefits of lightweight conversational control. In parallel, Basta et al.'s \emph{Bot Wars Evolved}~\cite{basta2025bot} explored adversarial self-play between LLM-based attacker and defender agents. By incorporating role-specific objectives and chain-of-thought prompting, their system sustained coherent, strategic dialogues that mirrored real scambaiting data.

Beyond engagement, GenAI is also being used for predictive defenses. ScamGPT-J~\cite{tan2024scamgpt}, for example, anticipates scammer utterances in real-time and alerts users based on conversational patterns. These developments highlight a growing shift from passive detection to active, generative defense mechanisms in combating online scams.

Together, these systems illustrate a growing trend toward leveraging  GenAI for proactive engagement with adversaries. However, most prior studies are limited to controlled simulations, short-term deployments, or small-scale datasets. In contrast, our work presents the first evaluation of a large-scale, real-world deployment of a generative scambaiting system. By analyzing high-volume and longitudinal interaction data with actual scammers, we provide a more comprehensive understanding of how such systems perform in practice and contribute to threat intelligence collection.

\subsection{Metrics for Evaluating Scambaiting Systems}

Prior work on scambaiting systems has explored a diverse set of evaluation methodologies, spanning structural metrics (e.g., engagement length, call duration), functional outcomes (e.g., information disclosure, goal fulfillment), heuristic measures (e.g., persona coherence, response diversity), and subjective assessments (e.g., realism, role fidelity, convincingness). \emph{Re:Scam}~\cite{netsafe2024,netsafe2024rescam} focused on scammer time wasted, engagement realism, and persona diversity. \emph{Puppeteer}~\cite{charnsethikul2025puppeteer} measured information disclosure rate, persona coherence, and convincingness. \emph{Bot Wars Evolved}~\cite{basta2025bot} emphasized engagement length, goal fulfillment, and conversational realism. Table~\ref{tab:evaluation-criteria} summarizes these metrics across systems.

Our research extends this landscape by introducing a more comprehensive and operationally grounded evaluation framework. Specifically, we extend beyond isolated disclosures and human-rated assessments by introducing longitudinal metrics such as \emph{Information Disclosure Speed (IDS)} and \emph{Takeoff Ratio}, which quantify not only whether but also how quickly and reliably engagements yield intelligence. We further evaluate message quality through \emph{Human Acceptance Rate (HAR)} and \emph{Edit Distance}, capturing real-world usability in human-in-the-loop workflows, an area underexplored in earlier works. These additions enable a richer, end-to-end understanding of scambaiting system performance in practical, large-scale deployments.

% Our research uniquely extends this evaluation paradigm by introducing quantitative intelligence-centric metrics, such as the \emph{Intelligence Acquisition Ratio} and \emph{Intelligence Yield Rate}, systematically assessing the practical effectiveness of deployed adversarial dialogue systems.

\section{System Overview}\label{sec:system}
Our analysis is based on data collected from a real-world LLM-powered scambaiting platform. The system operates using a single-prompt architecture, in which a prompt is submitted to a ChatGPT-based model to generate the next best response in an ongoing engagement with a scammer.

This platform includes a user interface that displays threaded email exchanges between the scammer and the defender agent (the fake victim persona). When a scammer's message requires a reply, the system suggests a response in a textbox. A human operator then reviews the suggestion, optionally edits it, and approves it for sending. This human-in-the-loop design balances the creative flexibility of the LLM with oversight to maintain coherence, appropriateness, and operational safety. The system supports multiple email accounts and predefined personas that impersonate different types of scam targets. These personas vary in tone, background, and behavioral style, enhancing the realism of the interactions.

Since this system is deployed in a real-world setting, both the prompt templates and response strategies have evolved organically throughout the five-month data collection period. In some cases, human experts intervened to manually craft or refine responses, especially in high-stakes or ambiguous exchanges. Figure~\ref{fig:system-architecture} illustrates the overall system architecture, including the message flow, prompt generation, LLM response handling, and human review loop.

\section{Dataset} \label{sec:dataset}

\begin{table*}[h]
\centering
\caption{Summary of message- and engagement-level metrics by system mode.}
\label{tab:message_and_thread_stats}
\begin{tabular}{lrrr}
\toprule
\textbf {Metric} & \textbf{LLM-Only (Mode~I)} & \textbf{LLM + HITL (Mode~II)} & \textbf{Overall} \\
\midrule
\multicolumn{4}{l}{\textbf{Message-level Statistics}} \\
\midrule
Messages                      & \messagesmodeone & \messagesmodetwo & \totalmessages \\
Attacker Messages             & \attackermsgsone & \attackermsgstwo & 9{,}320 \\% (optional: \newcommand{\totalattackermsgs}{9{,}320})
Defender Messages             & \defendermsgsone & \defendermsgstwo & 9{,}477 \\ % (optional: \newcommand{\totaldefendermsgs}{9{,}477})
Time Span (days)              & \durationmodeone & \durationmodetwo & 154\\ % (optional: \newcommand{\totalduration}{154})
Mean Messages per Day         & \msgsperdaymodeone & \msgsperdaymodetwo & 134.1 \\% (optional: \newcommand{\meanmsgsperday}{134.1})
Mean Msgs per Engagement    & \msgsperconvone & \msgsperconvtwo & 7.4 \\% (optional: \newcommand{\msgsperconvoverall}{7.4})
Mean Msg Length (chars)       & \avgmsgcharone & \avgmsgchartwo & \avgmsgcharall \\
Special Content Messages      & 12 & 2 & \totalspecialcontentmessages \\
\midrule
\multicolumn{4}{l}{\textbf{Engagement-level Statistics}} \\
\midrule
Seeded Engagement          & \convsone & \convstwo & \allthreads \\
Matured Engagements (\%)         &  \threadsbyvalidmodeone(\threadsbyvalidmodeonepct) & \threadsbyvalidmodetwo (\threadsbyvalidmodetwopct)     & \respondedthreads\ (\takeoff) \\
Mean Message Count         & \meanlengthone & \meanlengthtwo & \meanthreadlength \\% (optional: \newcommand{\meanthreadlengthoverall}{12.1})
Mean Duration (days) & \avgdurationone & \avgdurationtwo & \avgthreadduration \\ % (optional: \newcommand{\meandurationoverall}{11.04})
\bottomrule
\end{tabular}
\end{table*}

\subsection{Data Collection} We acquired a dump of the system’s message database under a confidentiality agreement with the platform administrators. The dataset includes all messages exchanged between scammers and ten distinct identities, referred to as defender agents, along with accompanying metadata such as timestamps, sender and recipient email addresses, message subject lines, and flags indicating whether a scammer’s message contains financial account information (e.g., bank details). The database also contained AI-generated response suggestions that were presented to human operators for engaging with scammers.

The dataset reflects production data from the deployed system across several distinct operational phases, all utilizing the same user interface (UI). The dataset used in this study includes data from two distinct operational modes of the system:

\begin{itemize}
    \item \textbf{Mode I: LLM-Only.} In this mode, the system autonomously generated all responses to scammers. The messages were sent to scammers \emph{without} any edit. However, for safety reasons, a human operator was still required to manually review messages before sending. This configuration was active for 120 days.
    
    \item \textbf{Mode II: LLM with Human-in-the-Loop (LLM + HITL).} In this mode, human operators reviewed and optionally edited LLM-generated messages before they were sent. This configuration was used for 34 days.
\end{itemize}

These two modes are referenced throughout the paper to distinguish between the LLM-only and LLM-with-HITL configurations. Our analysis covers the full 5-month-and-5-day deployment, examining both modes collectively and separately where appropriate. The LLM-only mode serves as the baseline for assessing the impact of human intervention, including its effect on system performance and the acceptance rate of LLM-generated messages.

\subsection{Overall Statistics} 

\noindent\textbf{Overall Message Statistics.} We analyzed a total of \totalmessages\ messages exchanged between scammers and defender agents on the scambaiting platform, spanning two system modes. Mode~I (LLM-only) contributed \messagesmodeone\ messages (69.2\%), while Mode~II (LLM + HITL) contributed \messagesmodetwo\ messages (30.8\%).

Engagements in Mode~I spanned \durationmodeone\ days, while those in Mode~II occurred over just \durationmodetwo\ days. Despite the shorter duration, Mode~II exhibited higher messaging activity, averaging \msgsperdaymodetwo\ messages per day compared to \msgsperdaymodeone\ in Mode~I. However, the average number of messages per engagements remained similar, with \msgsperconvone\ in Mode~I and \msgsperconvtwo\ in Mode~II.

Overall, attacker messages were substantially longer than those from defenders, with average lengths of \avgattackerchar\ and \avgdefenderchar\ characters, respectively. This asymmetry reflects the typical structure of scammer communication, which often relies on persuasive narratives or detailed instructions, while defender responses tend to be shorter and more reactive.

Special content, such as attachments, was exceedingly rare. Only \totalspecialcontentmessages\ messages across the entire dataset (\specialcontentpercent) included non-text formats: four PDFs, thirteen images, and one RTF file. This supports the observation that scammers in this dataset primarily relied on text-based persuasion rather than multimedia payloads.

\noindent\textbf{Overall Engagement Statistics.} Throughout the paper, we use the following terminology to describe the different types of engagements in our dataset:

\begin{itemize}
    \item \textbf{Seeded Engagements:} Refers to any initiated conversation or thread, including those that did not receive any response from scammers. Also referred to as \textit{All Engagements}.
    
    \item \textbf{Matured Engagements:} A conversation that received at least one response from a scammer.
    
    \item \textbf{Successful Engagements:} A matured conversation in which the scammer disclosed sensitive information, such as mule account details.
\end{itemize}
The dataset comprises~\allthreads\ unique targeted scamming email addresses. In \singleresponsecount\ cases (\singleresponsepercent) scammer did not respond to the defender’s initial message. Further investigation revealed that approximately half of these unresponsive email addresses were inactive by the time we attempted engagement, despite having previously been used to distribute scam emails.

Out of the \allthreads{} seeded engagements, \respondedthreads{} (\takeoff{}) matured into active two-way exchanges involving participation from both the scammer and the defender. These matured engagements had an average length of \meanthreadlength\ messages (median: \medianthreadlength), ranging from \threadlengthmin\ to \threadlengthmax\ messages, with a standard deviation of \threadlengthstddev. The distribution of thread lengths showed that {\threadstwotofivepct} had 2--5 messages, {\threadssixtotenpct} had 6--10 messages, and \threadseleventotwenty\ reached 11--20 messages. Matured engagements exceeding 20 messages accounted for 19.6\%, including \threadshundredplus\ matured engagements with over 100 messages. The substantial effort involved highlights the necessity of automating scambaiting to scale threat intelligence gathering.

The durations of matured engagements varied significantly, with an average duration of \avgthreadduration\ days and a median of \medianthreadduration\ days. The longest exchange spanned \longestduration\ days and contained \longestdurationmsgcount\ messages, while the most message-dense thread featured \longestthreadlen\ messages over \longestthreaddur\ days. Mode-specific analysis revealed that Mode~I matured engagements were generally longer and lasted longer than Mode~II engagements. Mode~I produced \validthreadsone\ matured engagements, with an average length of \meanlengthone\ messages and an average duration of \avgdurationone\ days, whereas Mode~II yielded \validthreadstwo\ matured engagements, averaging \meanlengthtwo\ messages in length and \avgdurationtwo\ days in duration. Detailed statistics on engagements are presented in Table~\ref{tab:message_and_thread_stats}.

\subsection{Legality and Ethics of Data} Despite its defensive potential, scambaiting introduces important ethical and legal considerations. Active defense framework~\cite{blair2016into} provides essential guidelines to ensure that proactive security measures, such as deceptive engagements and honeypot interactions, remain both legally compliant and ethically sound. Lundie et al.~\cite{lundie2024enterprise} stress the importance of clear operational boundaries, avoiding entrapment, safeguarding privacy, and maintaining transparency to legitimize scambaiting practices. In this work, all data was handled carefully by an enterprise entity operating fully within the bounds of U.S. law and industry best practices. The email addresses targeted by the enterprise entity were collected through spam honeypots and other scambaiting platforms, ensuring with high confidence that they belonged to known scammers. The collected threat intelligence was responsibly and promptly shared with financial institutions (e.g., FS-ISAC) and appropriate legal authorities to enhance user protection. Ethical handling of data and respect for legal standards were central to every stage of data collection and analysis.

\section{Evaluation Metrics for Scambaiting Effectiveness} \label{sec:evaluation}

%To enable meaningful benchmarking and research progress in automated scambaiting, we define a set of evaluation metrics that capture both the engagement success with scammers and the quality of generated messages. Several of these metrics are inspired by prior systems such as \textit{Puppeteer}~\cite{charnsethikul2025puppeteer}, \textit{BotWar}~\cite{basta2025bot}, \textit{Re:Scam}~\cite{netsafe2024}, and \textit{Moelley}~\cite{molloy2023scamming}. Others are newly proposed to reflect challenges observed in real-world deployments.
To enable meaningful benchmarking and research progress in automated scambaiting, we define a set of evaluation metrics that capture three complementary dimensions: (A) \emph{Disclosure Success}, meaning overall success in extracting sensitive information, (B) \emph{Message Generation
Quality}, and (C) \emph{Engagement Dynamics} meaning quality of engagement dynamics. Several of these metrics are inspired by prior systems such as \textit{Puppeteer}~\cite{charnsethikul2025puppeteer}, \textit{BotWar}~\cite{basta2025bot}, and \textit{Re:Scam}~\cite{netsafe2024}, while others are newly proposed to reflect challenges observed in real-world deployments. Table~\ref{tab:evaluation-metrics} presents the evaluation metrics applied in our dataset analysis.\\
For rest of the paper, if \( A \) is a finite set, the \emph{cardinality} of \( A \), denoted \( |A| \), is the number of elements in \( A \).

\begin{table*}[h]
\centering
\caption{Our Proposed Metrics for Scambaiting Effectiveness}
\label{tab:evaluation-metrics}
\begin{tabular}{@{}p{3.2cm}p{3.5cm}p{5.8cm}@{}}
\toprule
\textbf{Category} & \textbf{Metric} & \textbf{Description} \\
\midrule

\multirow{2}{=}{Disclosure Success} 
& Information Disclosure Rate (IDR) & Proportion of successful engagements that result in the extraction of sensitive information from scammers. \\
& Information Disclosure Speed (IDS) & Measures how quickly disclosures occur, in terms of message turns or elapsed time. \\

\midrule

\multirow{3}{=}{Message Generation Quality}
& Human Acceptance Rate (HAR) & Proportion of LLM-generated messages accepted by human operators without edits. \\
& Average Edit Distance & Mean Levenshtein distance between original LLM suggestions and final human-edited versions. \\
& Message Freshness & Proportion of novel $n$-grams in a message compared to the sender’s prior history, indicating linguistic diversity. \\

\midrule

\multirow{3}{=}{Engagement Dynamics}
& Takeoff Ratio & Percentage of matured engagements that receive at least one scammer response. \\
& Engagement Endurance & Measures engagement depth through average number of message turns and duration (in days). \\
& Response Invocation & Average time delay between defender messages and scammer replies, capturing scammer reactivity. \\

\bottomrule
\end{tabular}
\end{table*}

\subsection{Base Metrics: Measuring Disclosure Success}

These metrics focus on the primary objective of scambaiting systems: extracting sensitive information from scammers.

\noindent\textbf{Information Disclosure Rate (IDR).} This metric measures how effective the baiting agents are at eliciting sensitive information from scammers, such as bank account numbers or cryptocurrency wallet addresses.

\[
\text{IDR} = \frac{|\text{Successful engagements}|}{|\text{Total Engagements}|} \times 100
\]

This formulation mirrors intelligence-extraction metrics used in \textit{Puppeteer}.

\noindent\textbf{Information Disclosure Speed (IDS).}
This metric quantifies how quickly sensitive information is obtained from scammers. It captures both the number of engagements and the number of message turns required to reach a disclosure event.

%\[
%\text{IDS}_{\text{threads}} = %\frac{\text{\# Threads with disclosure}}%{\text{\# Total threads}}
%\]
\[
\text{IDS}_{\text{turns}} = \frac{1}{D} \sum_{i=1}^{D} \text{Turns until disclosure}_i
\]

where \( D \) is the number of successful engagements in which a disclosure occurred. A lower IDS indicates faster extraction of valuable information, and can reflect both the persuasiveness of the baiting agent and the scammer's willingness to escalate.

\subsection{Message Generation Quality}
These metrics assess the quality, novelty, and human usability of LLM-generated messages.

\noindent\textbf{Human Acceptance Rate (HAR).}
In seeded engagements involving human-in-the-loop (HITL) interventions, HAR quantifies the proportion of model-generated messages that are accepted by human operators without modification.

\[
\text{HAR} = \frac{|\text{Unedited messages}|}{|\text{Messages reviewed}|} \times 100
\]

To more finely assess message quality and required effort, we additionally compute the average Levenshtein distance between the original LLM output and its final edited version:

\[
\text{Avg. Edit Distance} = \frac{1}{K} \sum_{k=1}^{K} \text{Levenshtein}(m_k, m_k')
\]

where \( m_k \) is the original message and \( m_k' \) is the human-edited version, and K is the total edited messages. This complements HAR by capturing the extent, not just the frequency, of required human corrections. A related concept of edit ratio appears in \textit{Puppeteer}.

%\textcolor{red}{Word count, Levensten, sentiment analysis, tone, question vs statement, dialogue act}

\noindent\textbf{Message Freshness.} 
We define \textit{Message Freshness} as the proportion of novel $n$-grams in a message relative to all $n$-grams seen so far from the same sender type (attacker or defender). For a message $m_i$ by sender $s \in \{\text{attacker}, \text{defender}\}$, let $\text{ngrams}(m_i, n)$ denote the set of all contiguous $n$-grams in $m_i$, and let $\mathcal{H}_s^{(i)}$ denote the union of all $n$-grams used in previous messages by $s$ up to but not including $m_i$.

Then the message freshness score is given by:

\[
\text{Freshness}(m_i, n) = \frac{|\text{ngrams}(m_i, n) - \mathcal{H}_s^{(i)}|}{|\text{ngrams}(m_i, n)|}
\]

This value ranges from 0 (fully repetitive) to 1 (completely novel). We compute this metric across different $n$-gram sizes (e.g., $n = 2, 3, 4$) to assess both lexical and structural diversity.

%\textcolor{red}{See how the attacker and defender messages are fresh, how much repetition we see, etc}

%\noindent\textbf{Effectiveness of trigger message.} 
%\textcolor{red}{Given if the defender asked atatcker to provide bank account, how successfull it was} 

\subsection{Engagement Dynamics}
These metrics capture how well the system sustains scammer engagement, encouraging prolonged interactions that can lead to disclosures.

\noindent\textbf{Takeoff Ratio.} This metric captures the fraction of seeded engagements that evolve into matured engagements, defined as receiving at least one response from a scammer.

\[
\text{Takeoff Ratio} = \frac{|\text{Matured engagements}|}{|\text{Seeded engagements}|} \times 100
\]

%\begin{itemize}
%    \item \textcolor{red}{content size/freshness/fluctuation, correlation between trun number and length of conversation, etc}
%    \item  \textcolor{red}{does the attacker get faster or slower overtime, or they become faster before they disclose their info ...}
%\end{itemize}

\noindent\textbf{Engagement Endurance.}
Engagement Endurance reflects the depth of the scammer-baiter interaction, and can be quantified either by the number of turns or by elapsed time.

\[
\text{Average Turns} = \frac{1}{N} \sum_{i=1}^{N} \text{Turns}_i
\]
\[
\text{Average Duration} = \frac{1}{N} \sum_{i=1}^{N} \text{Duration}_i
\]

Both metrics have been used in prior works, including \textit{BotWar} and \textit{Re:Scam}, to assess sustained scammer engagement.

\noindent\textbf{Response Invocation.}
This metric captures the scammer's reactivity by measuring the average response time between a bait message and the scammer's reply. Prompt responses may indicate greater scammer interest or urgency.

\[
\text{Average Response Time} = \frac{1}{M} \sum_{j=1}^{M} \left( t^{\text{reply}}_j - t^{\text{bait}}_j \right)
\]

%\noindent\textbf{Survival Curve.} \textcolor{red}{TODO}

\section{Analysis of System Effectiveness} \label{sec:analysis}

We analyze our deployment results using the metrics defined in Section~\ref{sec:evaluation}. 
The findings are grouped into three categories: 
(A) \textit{Base Metrics}, 
(B) \textit{Engagement Quality}, and 
(C) \textit{Message Generation Quality}.

\subsection{Base Metrics: Measuring Disclosure Success}

\subsubsection{\textbf{Information Disclosure Rate (IDR)}}

Out of \allthreads{} seeded engagements, \IDRsuccessthreads{} led to mule account disclosures, corresponding to an overall Information Disclosure Rate (IDR) of \IDRoverall{}. When broken down by mode, Mode~I yielded an IDR of \IDRmodeone{}, and Mode~II recorded \IDRmodetwo{}.

Restricting the analysis to matured engagements, those involving more than one message exchange, the IDR rises notably to \IDRconvonly{}. Within this subset, Mode~I produced an IDR of \IDRmodeoneconv{}, while Mode~II achieved a slightly higher rate of \IDRmodetwoconv{}. Table~\ref{tab:idr_summary} summarizes IDR values across modes and thread types.

\begin{table}[h]
\centering
\caption{Summary of Information Disclosure Rate (IDR) segmented by operational mode and engagement type. ``Matured Engagements'' denote engagements in which the scammer responded at least once.}
\begin{tabular}{lcc}
\toprule
\textbf{Thread Type} & \textbf{Mode} & \textbf{IDR (\%)} \\
\midrule
All Engagements          & All Modes & \IDRoverall{} \\
                           & Mode~I    & \IDRmodeone{} \\
                           & Mode~II   & \IDRmodetwo{} \\
\midrule
Matured Engagements    & All Modes & \IDRconvonly{} \\
                           & Mode~I    & \IDRmodeoneconv{} \\
                           & Mode~II   & \IDRmodetwoconv{} \\
\bottomrule
\end{tabular}
\label{tab:idr_summary}
\end{table}

\subsubsection{\textbf{Information Disclosure Speed (IDS)}}

Among the \IDRsuccessthreads\ successful engagements, disclosures occurred after an average of 10.3 message turns, with a median of 8.0 turns and a range from 2 to 60. In terms of time, the average time to disclosure was 7.4 days, while the median was 3.3 days, with the slowest case taking over 120 days. Categorizing by speed, 12.9\% of disclosures occurred within 2--3 turns (Very Fast), 33.0\% in 4--7 turns (Fast), and 35.4\% in 8--15 turns (Medium), while only 3.4\% required more than 30 turns. Notably, 30.3\% of disclosures were completed within 5 turns, and 21.5\% occurred within 24 hours. 

\begin{figure}[t!]
    \centering
    \includegraphics[width=0.45\textwidth]{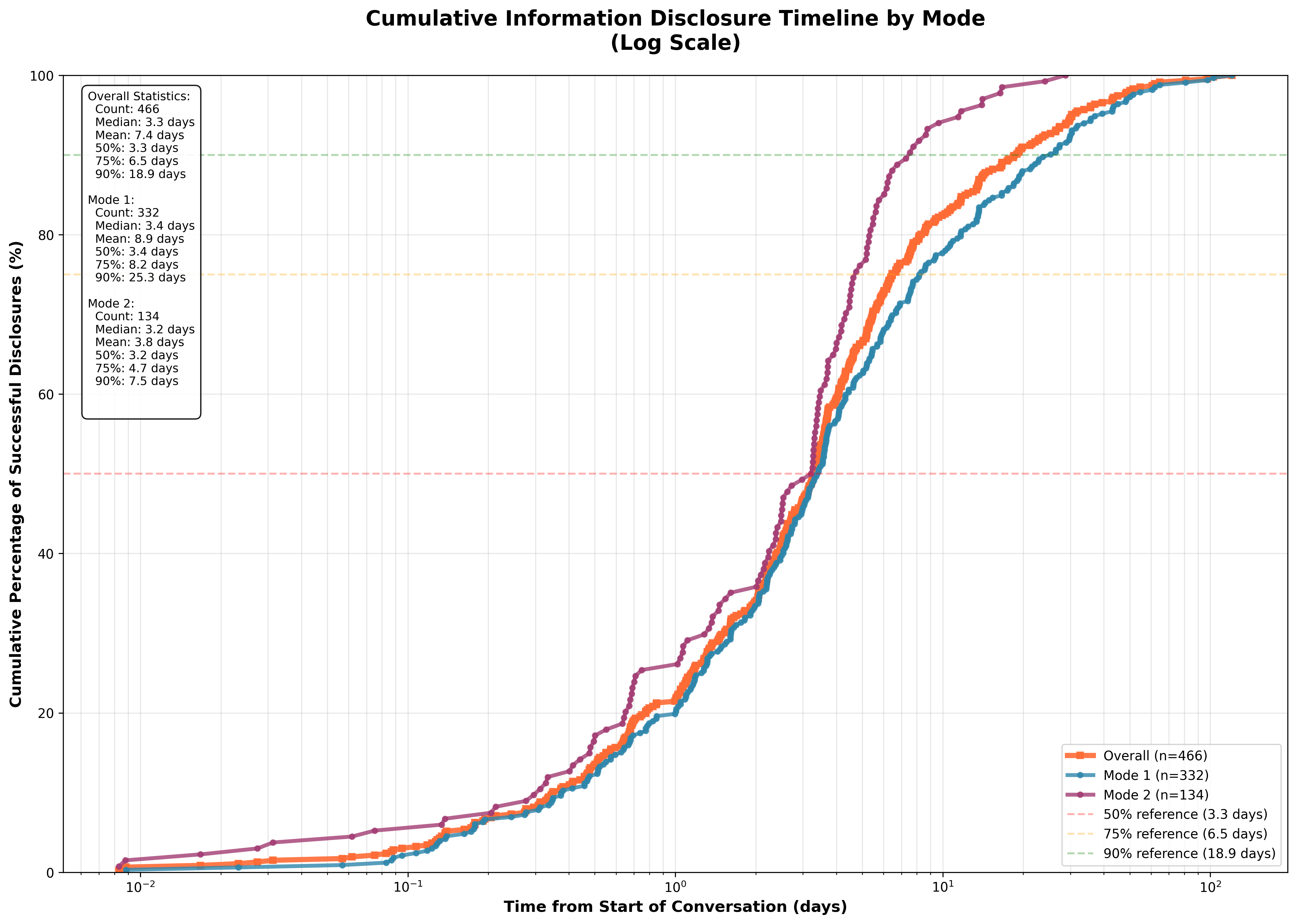}
    \caption{Information Disclosure Speed: Cumulative timeline of successful information disclosures by mode (log scale). The plot shows the cumulative percentage of disclosures over time since the start of each engagement, separated by Mode~I and Mode~II. Disclosures occur rapidly in both modes, with 50\% of successful cases revealed within 3.3 days overall. Mode~II achieves faster disclosures, reaching the 90\textsuperscript{th} percentile at just 7.5 days, compared to 25.3 days in Mode~I.}
    \label{fig:ids_cumulative_timeline}
\end{figure}

Figure~\ref{fig:ids_cumulative_timeline} illustrates the cumulative distribution of successful information disclosures over time, using a logarithmic scale. The plot distinguishes between disclosures across all engagements, as well as those in Mode~I and Mode~II. Overall, 50\% of disclosures occurred within 3.3 days of engagement initiation, and 90\% were obtained within 18.9 days. Mode~II demonstrates a faster disclosure timeline, with a median of 3.2 days and 90\% of disclosures completed by 7.5 days. In contrast, Mode~I shows a broader spread, with a 90\textsuperscript{th} percentile of 25.3 days. These findings suggest that Mode~II facilitates quicker extraction of mule account information, likely due to its more aggressive or adaptive interaction strategy. 

Comparing modes, Mode~II (HITL) outperformed the fully automated Mode~I in both conversational and temporal efficiency. Mode~II achieved disclosures in fewer turns on average (9.1 vs.\ 10.8) and in less time (3.8 vs.\ 8.9 days).

\iffalse
\begin{table*}[h]
\centering
\caption{Detailed Summary of Information Disclosure Speed (IDS)}
\begin{tabular}{llccc}
\toprule
\textbf{Category} & \textbf{Subset} & \textbf{Engagements (\%)} & \textbf{Avg Turns} & \textbf{Avg Time (days)} \\
\midrule
\multirow{3}{*}{Overall} 
  & All Successful Engagements       & \IDRsuccessthreads\ (100.0\%) & 10.3 & 7.4 \\
  & $\leq$5 Turns                & 141 (27.7\%)  & 3.3  & 3.4 \\
  & $\leq$24 Hours               & 100 (19.6\%)  & 4.3  & 0.4 \\
\midrule
\multirow{5}{*}{Speed Category} 
  & Very Fast (1--3 turns)       & 60 (12.9\%)   & 2.2  & 3.2 \\
  & Fast (4--7 turns)            & 154 (33.0\%)  & 5.2  & 4.3 \\
  & Medium (8--15 turns)         & 165 (35.4\%)  & 10.8 & 9.0 \\
  & Slow (16--30 turns)          & 71 (15.2\%)   & 21.2 & 11.4 \\
  & Very Slow ($>$30 turns)      & 16 (3.4\%)    & 37.1 & 19.1 \\
\midrule
\multirow{2}{*}{By Mode} 
  & Mode~I                       & 332 (71.5\%)  & 10.8 & 8.9 \\
  & Mode~II                      & 134 (28.5\%)  & 9.1  & 3.8 \\
\bottomrule
\end{tabular}
\label{tab:ids_detailed_summary}
\end{table*}

\fi

\subsection{Message Generation Quality}

\subsubsection{\textbf{Human Acceptance Rate (HAR)}}

We utilize human feedback provided in Mode~II as a proxy for evaluating the quality of AI-generated messages. A suggested response is considered a \emph{perfect acceptance} if it is sent to the scammer without any modification. For non-accepted responses, we quantify the extent of human intervention by measuring the degree of edits applied to the AI-generated message prior to sending.

For each of the \defendermsgstwo{} defender responses in Mode~II, an AI-generated reply was proposed. Of these, \harperfectmatch{} were accepted without any edits, yielding a Human Acceptance Rate (HAR) of \humanacceptancerate{}. The remaining responses were modified to varying extents, with an average edit distance of \avglevdistance\ characters (median: 0.0; range: 0--535; SD: 55.8). This pattern indicates a bimodal distribution: a large portion of suggestions were used verbatim, while others required substantial revision.

Acceptance rates varied by degree of modification. Minor edits ($\leq$5 characters) accounted for 35.5\% of all suggestions, followed by moderate edits (6--20 characters, 10.1\%), significant changes ($>$50 characters, 10.5\%), and major edits (21--50 characters, 2.1\%). This shows that a substantial portion of AI outputs were either trusted outright or adjusted lightly to fit the defender's intent.

Message length had a strong effect on HAR. Among short messages ($\leq$50 characters), only 0.41\% were accepted verbatim, and even minor edits averaged 11.6 characters, suggesting these shorter suggestions were less useful or harder to align with human preferences. In contrast, medium-length messages (51--200 characters) had the highest acceptance rate at 72.83\%, with a comparable average edit distance (24.0 characters). Long (201--500 characters) and very long messages ($>$500 characters) saw lower HARs of 49.25\% and 2.08\%, respectively, suggesting diminishing returns for lengthy AI-generated content.

Notably, acceptance patterns varied depending on the success of the engagement. In successful engagements (those resulting in mule account disclosure), 51.5\% of suggestions were perfectly accepted, compared to only 33.2\% in unsuccessful ones. Conversely, minor edits ($\leq$5 characters) were more common in unsuccessful engagements (46.9\%) than in successful ones (22.7\%). Successful interactions also saw higher rates of significant edits ($>$50 characters), suggesting a willingness to adapt AI outputs more creatively in high-stakes scenarios. These differences indicate that human defenders are more likely to accept AI suggestions directly when engagements are going well, and more likely to modify suggestions when engagement is challenging.

Overall, these findings highlight that AI assistance in Mode~II was both frequent and effective. While fewer than half of the suggestions were adopted without change, most edits were minor, and medium-length AI-generated replies were particularly likely to be trusted. Furthermore, the high HAR in successful engagements suggests that human acceptance of AI-generated messages may be an important proxy for conversational alignment and outcome quality.

\begin{figure}[ht]
    \centering
    \includegraphics[width=0.8\linewidth]{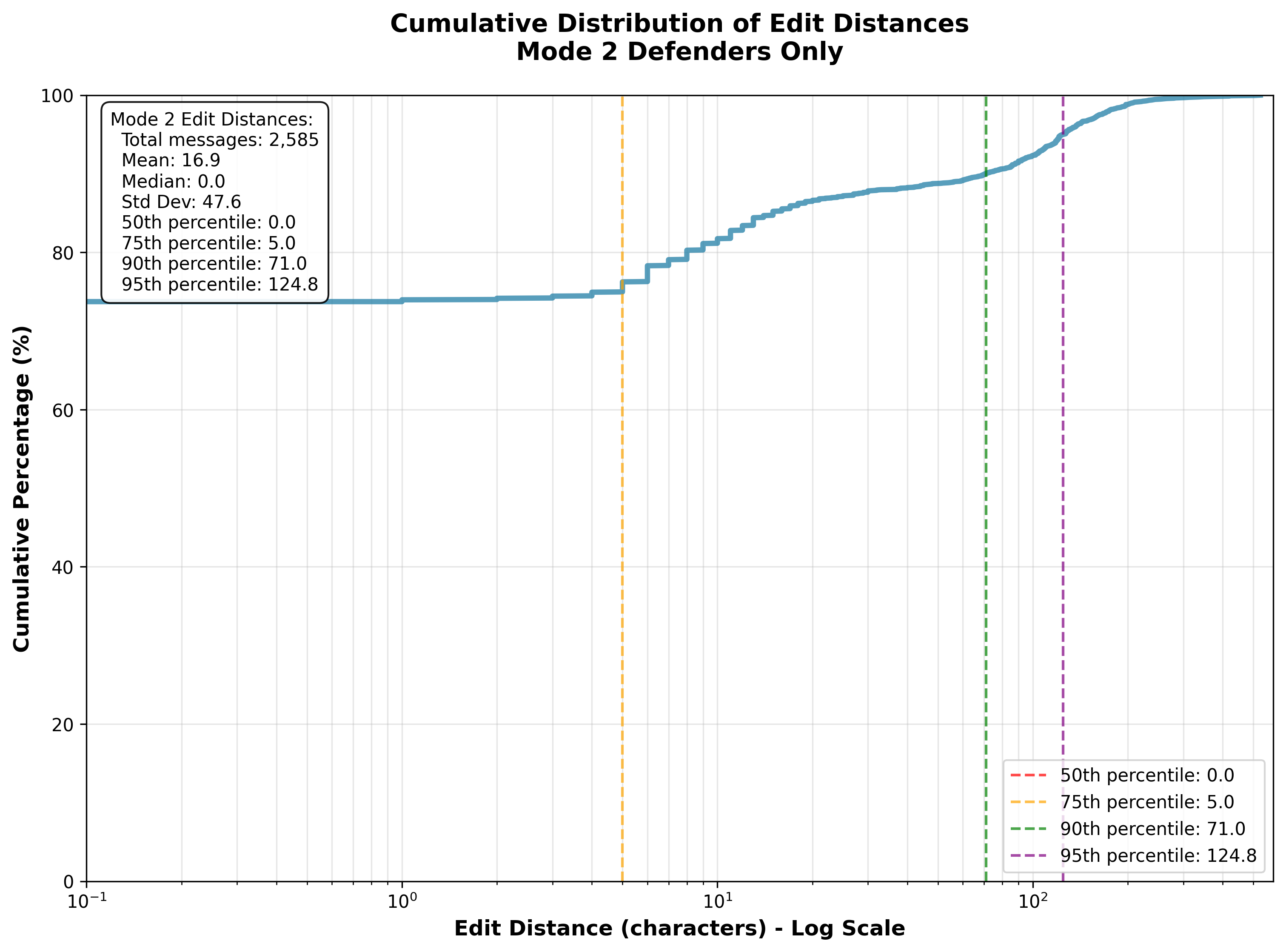}
    \caption{
Human Acceptance Rate: Cumulative distribution of edit distances for LLM-generated messages presented to human defenders. A total of \humanacceptancerate\ of the messages were accepted and sent without any modification. 
    }
    \label{fig:mode2-edit-distance}
\end{figure}

\subsubsection{\textbf{Message Freshness}}  \label{sec:message-freshness}

We analyzed message freshness for \totaldbmessages\  messages using $n$-gram sizes from 2 to 10. As $n$ increases, message freshness rises: 2-gram freshness averaged 0.158 (median 0.111), while 10-gram freshness reached 0.669 (median 0.955). This reflects high repetition in short phrases and greater diversity in longer expressions.

Attacker messages were consistently fresher than defender messages. For example, 4-gram freshness averaged 0.540 for attackers and 0.415 for defenders. Similarly, LLM-only Mode~I had higher freshness than HITL-assisted Mode~II, especially for small $n$ (e.g., 2-gram: 0.174 vs. 0.122). This suggests HITL intervention favors consistency over novelty.

Freshness distributions were heavily skewed at low $n$: nearly 47\% of messages had 2-gram freshness below 0.1. At 10-grams, over 55\% of messages scored above 0.9, highlighting the prevalence of reused short phrases versus more unique longer ones.

Frequent low-freshness $n$-grams included expressions like \textit{``bank account''}, \textit{``thank you''}, and \textit{``send the''}. For longer sequences ($n > 5$), repeated patterns in defender messages included \textit{``provide the bank account details where I should send''}, \textit{``could you please provide the bank account details''}, and \textit{``let me know where to send the money''}. These high-frequency constructions were used across many threads with minimal variation, suggesting templated responses optimized for efficiency but low in novelty. In contrast, high-freshness messages were typically short, idiosyncratic, or context-specific.

Finally, message freshness increased with position in the engagement. First-position messages had the lowest average 2-gram freshness (0.075), while third-position messages rose to 0.175. This suggests that earlier messages rely on standard scripts, while later turns involve more adaptive language.

\subsection{Engagement Dynamics}

\subsubsection{\textbf{Takeoff Ratio}}
The takeoff ratio quantifies the matured engagements, proportion of defender-initiated engagements that receive at least one scammer response, thus transitioning into an active engagement. Out of \allthreads\ total engagements, \respondedthreads\ received a scammer response, resulting in an overall takeoff ratio of \takeoff. The remaining engagements failed to take off.

Breaking this down by system configuration, Mode~I (LLM-only) initiated \threadsbyallmodeone engagements with a takeoff ratio of \threadsbyvalidmodeonepct, while Mode~II (HITL) achieved a slightly higher takeoff ratio of \threadsbyvalidmodetwopct\ across \threadsbyallmodetwo\ initiated engagements. It is important to note that HITL could not have edited the very first message. So this improvement is perhaps random. Further investigation showed that approximately 50\% of non-responses occurred because the scammer's email account was already inactive at the time of initial contact.

A key factor influencing takeoff was the length of the first message. Interestingly, failed takeoffs had significantly longer opening messages (average: 551.3 characters) compared to successful ones (average: 225.9 characters), with a difference of 325.4 characters. This pattern suggests that overly verbose or complex messages may discourage scammers from responding. The highest takeoff ratio (53.71\%) occurred in the ``long'' category (301--500 characters), while both very short ($\leq$50 characters) and very long ($>$500 characters) messages performed worse, indicating an optimal range for engagement-seeding messages.

We also analyzed takeoff behavior by day of the week. The highest takeoff rates were observed on Monday (52.66\%) and Wednesday (51.94\%), with consistent engagement across most weekdays. In contrast, Sunday (42.63\%) and Thursday (44.33\%) exhibited the lowest takeoff ratios, possibly reflecting lower scammer availability or responsiveness on those days. While these fluctuations are modest, they suggest that weekday outreach, particularly early in the week, may slightly increase the likelihood of successful engagement. Scammers need to enjoy their weekend after all.

In sum, the takeoff ratio analysis reveals that approximately half of defender-initiated engagements succeed in sparking scammer engagement, with HITL (Mode~II) offering a modest edge. Message brevity, timing, and careful phrasing, potentially guided by human review, appear to significantly improve the likelihood of triggering a scammer response. Day-of-week patterns add a further layer of insight for optimizing outreach strategy.

\begin{figure}[ht]
    \centering
    \includegraphics[width=0.95\linewidth]{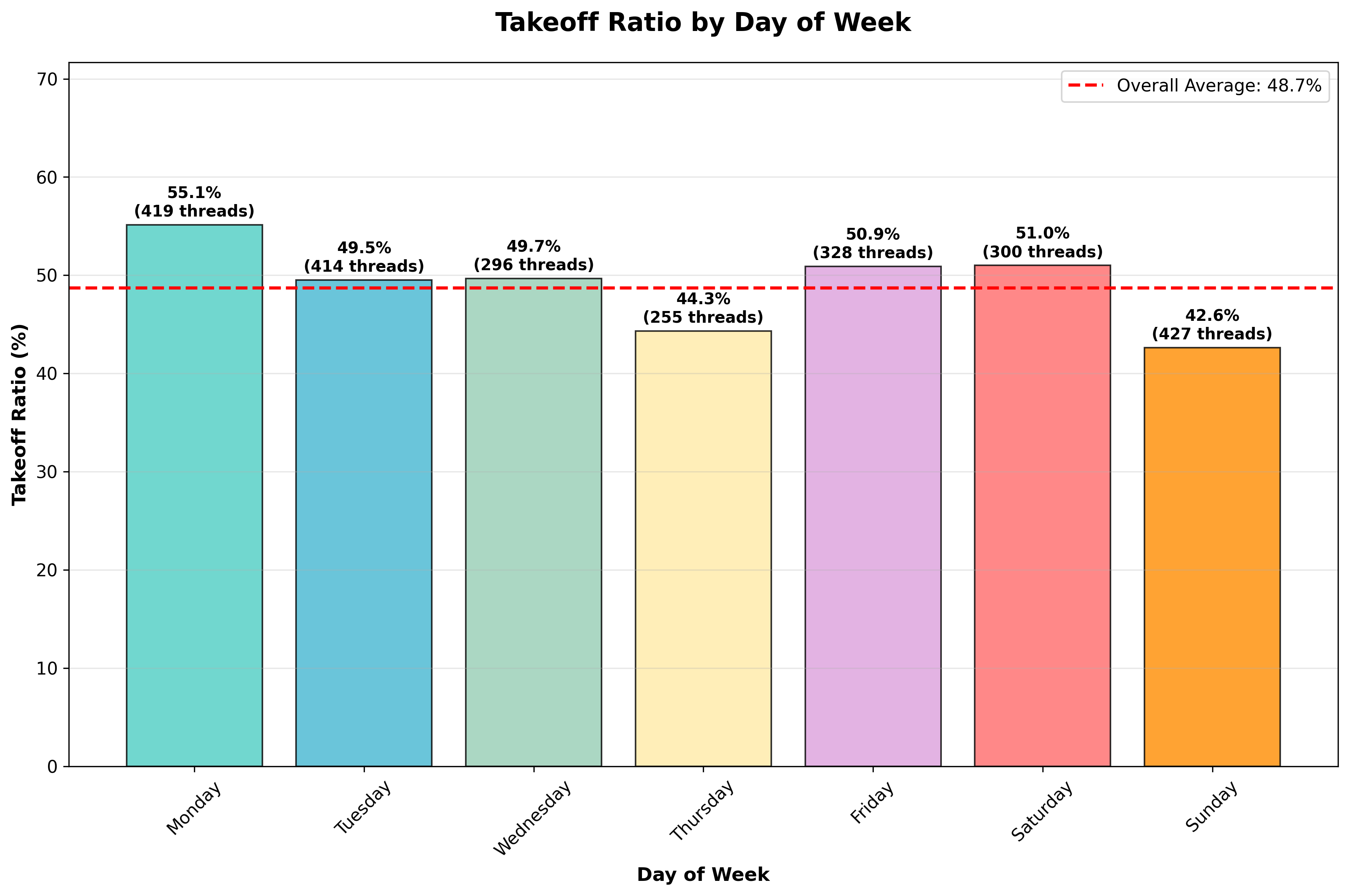}
    \caption{Takeoff Ratio: Monday exhibited the highest takeoff rate at 55.1\%, while Sunday had the lowest at 42.6\%. The overall average takeoff ratio across all days was \takeoff\ (indicated by the dashed red line). These results suggest moderate day-of-week variation in scammer responsiveness.}
    \label{fig:takeoff-by-day}
\end{figure}

\subsubsection{\textbf{Engagement Endurance}}
We examined engagement lengths across all \allthreads{} seeded engagement. Among them, \takeoff{} matured into two-way exchanges, with these matured engagements averaging 12.2 messages (median: 7.0), ranging from 2 to 206 messages.

Interaction durations varied widely as well. The average engagement lasted 261.1 hours (approximately 10.9 days), with a median of 116.6 hours (about 4.9 days), and a maximum duration of 3,336.2 hours (nearly 139 days). The standard deviation was 409.1 hours (17.0 days), highlighting the high variability of engagement timelines.

Classifying engagements by message count showed that 42.8\% of engagements were short (2--5 messages), 32.3\% were medium-length (6--15), 16.6\% were long (16--30), and 8.4\% were very long (31--50) or extreme (51+). While short engagements averaged 149.1 hours (about 6.2 days) in duration, very long and extreme ones extended to 501.5 hours (about 20.9 days) and 947.5 hours (about 39.5 days), respectively.

Grouping engagements by temporal duration revealed that 45.9\% were medium (1--7 days), 29.9\% were long (1--30 days), and 9.3\% extended beyond 30 days. A small fraction (0.4\%) concluded within one hour. Medium-duration engagements averaged 8.9 messages, while very long threads averaged 23.5 messages, reinforcing the link between time span and conversational depth.

A comparison of the two system modes revealed notable differences. Mode~I (LLM-only) engagements averaged 12.3 messages and 310.2 hours (about 12.9 days) in duration (median: 7.0 messages, 136.4 hours / 5.7 days), whereas Mode~II (LLM + HITL) averaged 11.7 messages and 128.4 hours (about 5.4 days) (median: 8.5 messages, 87.6 hours / 3.7 days). These results suggest that HITL setups support more efficient, time-bounded engagements while still sustaining conversational richness.

Success rates were also strongly tied to engagement length. Among the 1{,}599 multi-message matured engagements, 509 were successful in eliciting sensitive disclosures. These engagements averaged 23.4 messages over 354.0 hours (about 14.8 days) (median: 18.0 messages, 177.6 hours / 7.4 days). In contrast, the 1{,}090 unsuccessful engagements averaged only 6.9 messages and 217.7 hours (about 9.1 days) (median: 5.0 messages, 84.6 hours / 3.5 days), underscoring the importance of sustained engagement for scambaiting success.

In summary, longer and more persistent engagements, especially those involving human oversight, are more likely to result in successful outcomes. HITL-enhanced engagements achieve comparable message depth in significantly less time, highlighting their utility in efficient and effective threat intelligence operations.

\subsubsection{\textbf{Response Invocation}}
\label{sec:scammer-responsiveness}

We analyzed a total of \attackermsgsall\ scammer replies across \respondedthreads\ matured engagements to understand response behaviors. On average, scammers responded in 1{,}601.6 minutes (approximately 26.7 hours), with a median of 200.0 minutes (3.3 hours). Response times varied widely, ranging from 0.0 to 182{,}581.2 minutes (over 126 days), with a standard deviation of 7{,}484.3 minutes.

To characterize response latency, we grouped replies into six categories: immediate ($\leq$1 minute, 12.4\%), very fast (1--5 minutes, 1.3\%), fast (5--30 minutes, 14.7\%), medium (30--120 minutes, 14.9\%), slow (2--24 hours, 40.8\%), and very slow ($>$24 hours, 15.9\%).

\begin{figure}[t!]
    \centering
    \includegraphics[width=0.4\textwidth]{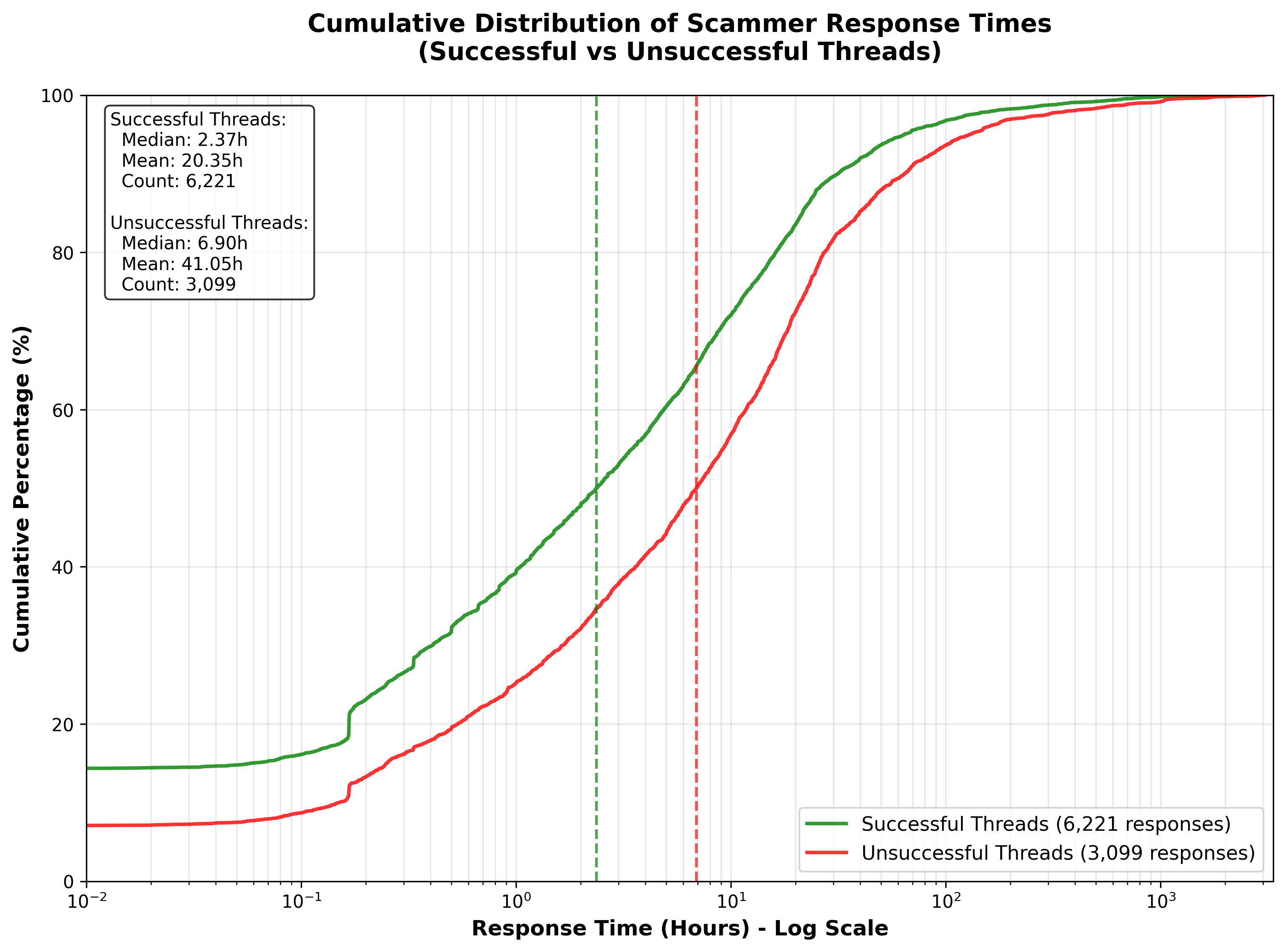}
    \caption{Response Invocation: Faster scammer responses are strongly associated with successful engagements. Response times in successful engagements are consistently shorter across all percentiles. On average, scammers respond within 20 hours in successful interactions, compared to nearly 41 hours in unsuccessful ones.}
\label{fig:scammer_response_time_cumulative}
\end{figure}

Response times were strongly linked to engagement outcomes. Figure~\ref{fig:scammer_response_time_cumulative} compares the cumulative distribution of scammer response times in successful versus unsuccessful engagements. Scammers in successful engagements responded significantly faster overall, with a median response time of 2.37 hours and a mean of 20.35 hours, compared to 6.90 hours (median) and 41.05 hours (mean) in unsuccessful engagements. This trend persists across percentiles: at the 50th percentile, successful responses were 4.54 hours faster, and at the 95th percentile, they were nearly 66 hours faster. These results suggest that quicker initial engagement by scammers is a strong indicator of eventual information disclosure.

System mode also influenced responsiveness. In Mode~I (LLM-only), average scammer response time was 1{,}446.1 minutes (24.1 hours), while Mode~II (LLM + HITL) exhibited longer average delays of 1{,}926.6 minutes (32.1 hours). However, within both modes, successful engagements were faster: in Mode~I, successful engagements averaged 1{,}083.4 minutes (18.1 hours) versus 2{,}262.4 minutes (37.7 hours) for unsuccessful ones.

The message position in the thread further impacted latency. The second scammer message exhibited the highest delay, averaging 2{,}833.4 minutes (47.2 hours). Response times declined in later positions, with eleventh-position replies averaging 931.7 minutes (15.5 hours). Longer engagements were also associated with quicker replies: engagements with over 30 messages had an average scammer response time of 944.3 minutes (15.7 hours), compared to 2{,}871.6 minutes (47.9 hours) in short engagements (2--5 messages).

Pairwise comparison of defender and scammer replies across 10{,}225 message pairs revealed a weak but positive correlation ($r = 0.083$), suggesting that faster defender responses may modestly encourage quicker scammer replies. Notably, defenders responded significantly faster on average (639.4 minutes; 10.7 hours) than scammers (1{,}593.9 minutes; 26.6 hours).

To better understand scammer disengagement behavior, we conducted a survival analysis based on the longest observed gaps in scammer responses following defender messages, referred to as the \textit{Death Pulse}. Among the \respondedthreadscount\ valid engagements, the longest observed scammer response gap was 126.8 days. On average, matured engagements experienced a maximum gap of 5.0 days, with a median of 1.1 days. These statistics suggest that while scammers often disengage early, a small fraction remain dormant for extended periods before occasionally resurfacing.

We computed survival rates by measuring the proportion of engagements that continued receiving scammer responses beyond specific time thresholds. The decay was steep: 52.1\% of engagements remained active after one day, 34.5\% after two days, and only 13.5\% remained responsive after one week. Beyond 30 days, fewer than 5\% of engagements remained active, and only 0.4\% survived past 100 days. The 95\% engagement cut-off defined as the time beyond which only 5\% of engagements remained alive, was measured at 28.0 days. This sharp drop-off underscores the urgency of early-stage engagement and intelligence extraction.

Comparing across system modes, Mode~I (LLM-only) engagements had a maximum scammer gap of 88.3 days, with an average of 3.9 days and a median of 1.0 day. Mode~II (LLM + HITL) engagements were slightly more persistent, with a higher maximum (126.8 days), longer average gap (6.8 days), and median of 1.1 days. Mode~II engagements also exhibited marginally higher short-term survival: 53.3\% persisted beyond one day, compared to 51.3\% for Mode~1. However, the difference narrowed over time; only 6.7\% of Mode~II and 3.0\% of Mode~I threads survived beyond 30 days.

Survival rates also varied based on the engagement outcome. Among the 33 successful engagements that resulted in mule account disclosures, the average scammer gap was 3.2 days (median: 1.5), compared to 5.1 days (median: 1.1) in the 1,252 unsuccessful ones. Furthermore, 63.6\% of successful engagements remained active past one day, but only 3.0\% survived beyond 14 days. These findings suggest that the most valuable threat intelligence is extracted early in the engagement, reinforcing the importance of optimizing early-phase interaction strategies.

\begin{figure}[h]
    \centering
    \includegraphics[width=0.9\linewidth]{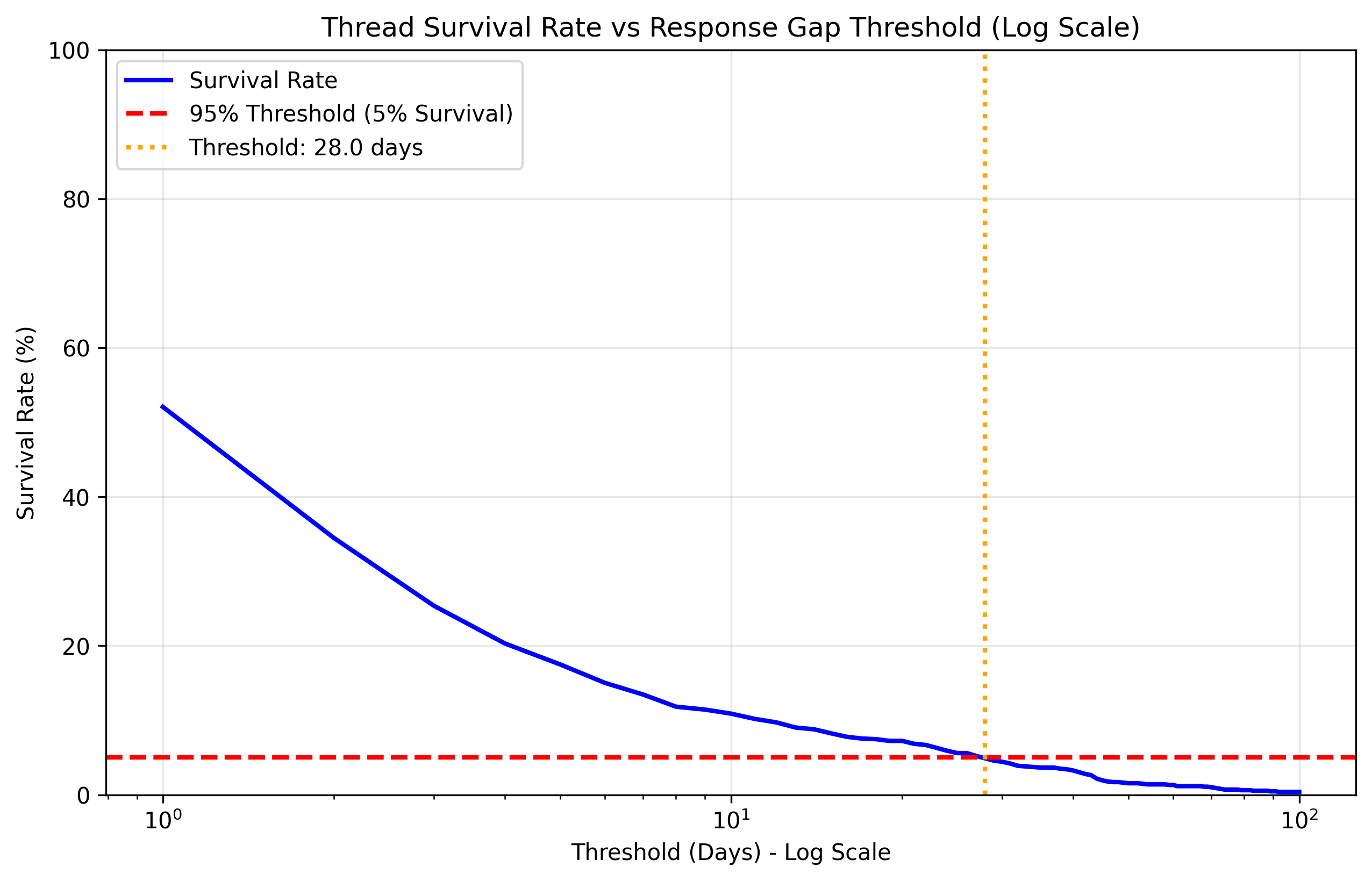}
    \caption{Survival rate of scammer responses over time. Engagements can be closed after 28 days of no response with 95\% confidence.}
    \label{fig:survival-plot}
\end{figure}

%\subsection{Metrics definition}
%Intelligence Acquisition Ratio
%Intelligence Yield Rate
%Success definitions (e.g., mule account disclosure)

%\subsection{Engagement statistics} Volume and duration (E1–E2)
%Latency and dynamics (E3–E4)
%Human intervention impact (E11, E14–E15)

\section{Lessons Learned and Design Insights} \label{sec:insights} 

%Highlevel Discussion
Our large-scale evaluation of an LLM-powered scambaiting deployment reveals consistent patterns in scammer–defender interactions. These findings clarify the strengths and limitations of autonomous conversational honeypots and highlight how targeted human oversight can amplify their impact. The seven insights that follow are derived from analyses based on the seven proposed metrics, organized into three categories, and provide an evidence-based foundation for designing more effective next-generation scambaiting systems.

\noindent \textbf{Insight I}: \textit{Sustained multi-turn engagement substantially increases the likelihood of eliciting sensitive intelligence, and limited human oversight improves this effect.} The IDR analysis shows that meaningful scammer intelligence depends on sustained engagement rather than initial outreach alone. Disclosures are uncommon in one-off exchanges, but once engagements progress beyond the first contact, the chances of uncovering mule account details increase significantly. This underscores persistence as a key driver of success: the longer we can keep scammers engaged, the greater the opportunity to gather actionable intelligence. Moreover, while autonomous systems can perform effectively on their own, introducing selective human oversight during active engagements provides a small but measurable benefit, improving the likelihood of disclosures.

\noindent \textbf{Insight II}: \textit{Timely and efficient engagement matters: most disclosures happen quickly once an engagement takes off, and human oversight accelerates this process.}
The IDS analysis reveals that once a scammer begins to engage, sensitive information is often disclosed relatively quickly, frequently within the first few days and, in many cases, within the first ten message exchanges. This highlights that early-stage dialogue management is critical: if momentum is built in the initial turns, the chance of success increases. Furthermore, engagements with limited human oversight progress to disclosures faster than fully autonomous exchanges, indicating that human-in-the-loop systems can guide interactions toward quicker and more efficient outcomes without requiring long, drawn-out exchanges.

\noindent \textbf{Insight III}: \textit{Human-in-the-loop feedback improves both the quality and impact of AI-generated messages, with acceptance patterns emerging as a strong proxy for conversational success.}
The Human Acceptance Rate (HAR) analysis shows that when human reviewers trust AI-generated responses without edits, the engagement is far more likely to lead to mule account disclosures. This highlights a dual role for humans: they act as quality filters, making small but impactful adjustments when necessary, and their acceptance behavior itself becomes a useful signal of conversational alignment. Medium-length, well-structured AI responses were particularly effective, while overly short or excessively long messages required more intervention. These findings suggest that human oversight not only strengthens AI-led engagements but also provides actionable indicators for where the system performs well and where future models can be optimized.

\noindent \textbf{Insight IV}: \textit{Message diversity increases over time, but templated responses dominate early turns.}
Our analysis of message freshness shows that initial responses, particularly those from defenders, tend to rely on highly repetitive templates, while diversity in phrasing grows as engagements progress. Although templates accelerate response generation, overuse reduces novelty and may risk early detection by scammers. This suggests that controlled variation and adaptive language generation could improve engagement realism without sacrificing efficiency.

\noindent \textbf{Insight V}: \textit{The ability to trigger scammer engagement (“takeoff”) depends critically on concise, well-timed initial outreach rather than system configuration.}
The takeoff ratio analysis reveals that only about half of initial outreach attempts progress into an active engagement, with long and verbose opening messages significantly reducing the chance of a scammer reply. Once the first response is secured, engagements typically develop into rich multi-turn interactions. Human-in-the-loop oversight offers only a marginal effect here, as the first message is unedited. Instead, timing and message style, particularly concise, natural-seeming openings, are the strongest determinants of whether a baiting attempt takes off. This underscores that the earliest moments of contact define whether intelligence-gathering opportunities will even begin.

\noindent \textbf{Insight VI}: \textit{Depth and persistence are critical: sustained engagements drive success, and human guidance makes them more efficient.}
The endurance analysis shows that successful intelligence gathering depends on keeping scammers engaged for longer engagements. Successful engagements that eventually produced mule account information were, on average, more than three times as long as unsuccessful ones. This pattern underscores that persistence, not just initial contact, is essential for eliciting disclosures. At the same time, introducing human oversight significantly shortens the time required to reach these outcomes: human-in-the-loop interactions achieve comparable conversational depth in roughly half the duration of fully automated exchanges. These findings highlight the dual importance of endurance and guided efficiency as complementary factors in successful scambaiting.

\noindent \textbf{Insight VII}: \textit{Scammer responsiveness is a signal of engagement quality and a predictor of eventual disclosure.}
Analysis of response latencies reveals that scammers who respond more quickly are significantly more likely to disclose mule account information. Faster response patterns, especially after the early stages of an engagement, indicate stronger engagement, while slow or inconsistent replies correlate with abandoned engagement. This suggests that latency itself can be a valuable operational signal: defenders can prioritize and adapt strategies in engagements where scammers are more responsive. Additionally, sustained back-and-forth accelerates over time, showing that once an engagement gains momentum, scammers tend to reply faster, making these interactions particularly promising for intelligence gathering.

\begin{table*}[h!]
\centering
\caption{Summary of Scambaiting Evaluation Results}
\label{tab:evaluation-summary}
\begin{tabular}{@{}p{3.2cm}p{7cm}p{3cm}@{}}
\toprule
\textbf{Category} & \textbf{Metric} & \textbf{Value} \\
\midrule
\multirow{2}{=}{Disclosure Success}
& Information Disclosure Rate (IDR) & \IDRconvonly \\
& Information Disclosure Speed (IDS) & 10.3 turns / 7.4 days \\
\midrule
\multirow{3}{=}{Message Generation Quality}
& Human Acceptance Rate (HAR) & \humanacceptancerate\ \\
& Average Edit Distance & \avglevdistance\ characters \\
& Message Freshness (4-gram) & 0.415 (defenders) \\
\midrule
\multirow{3}{=}{Engagement Dynamics}
& Takeoff Ratio & \takeoff\ \\
& Engagement Endurance (Avg Turns) & 12.2 turns / 10.9 days \\
& Response Invocation (Avg Response Time) & 26.7 hours \\
\bottomrule
\end{tabular}
\end{table*}

\section{Conclusion and Future Work} \label{sec:conclusion}

In this paper, we conducted a comprehensive analysis of an LLM-powered scambaiting system, focusing on its operational behavior, effectiveness, and conversational patterns during real-world interactions with scammers. Our evaluation was based on a dataset containing over \totaldbmessageshundreds\ messages across more than \allthreadshundreds\ engagement threads. We assessed system performance using eight distinct metrics, grouped into three overarching categories. Table~\ref{tab:evaluation-summary} presents the key performance metrics across these evaluation dimensions.

Our analysis revealed notable differences between fully automated and human-in-the-loop system configurations. While automated engagements demonstrated extensive long-tail interactions, the presence of human reviewers led to quicker and more efficient extraction of sensitive information, suggesting significant benefits in hybrid operational approaches. Furthermore, our engagement survival analysis provided insights into interaction longevity, crucial for optimizing timing and persistence strategies in scambaiting systems.
Our findings underline the potential of combining sophisticated conversational models with strategic human oversight, laying the groundwork for more robust, scalable, and adaptive active-defense solutions against evolving scam threats.

Despite these findings, several aspects warrant further investigation. Future work will delve deeper into the nuanced dialogue dynamics to explicitly capture state transitions and their predictive value in scammer behavior modeling. We aim to explore whether specific sequences or types of engagements can predict the successful elicitation of sensitive information, potentially improving the system's proactive engagement strategies.

Additionally, several exploratory analyses identified in our initial study offer promising avenues for future investigation. These include examining the qualitative impact of human intervention, specifically, how varying levels and types of human editing affect engagement outcomes, and analyzing the chronology of scammer intents and their correlation with specific scam types. Further, we aim to identify conversational breakpoints or exit signals that may precede information disclosures or attacker disengagement. Latency dynamics across different roles and scam categories also warrant investigation to better understand the role of urgency and timing in scam tactics.

Finally, incorporating automated and expert-based content evaluations alongside user or mission satisfaction metrics will provide holistic assessments of conversational quality. Such analyses will be instrumental in advancing our understanding of automated dialogue efficacy.

\bibliographystyle{IEEEtran}
\bibliography{main}

\end{document}